\documentstyle[12pt]{article}
\newcommand{\be}{\begin{equation}}
\newcommand{\ee}{\end{equation}}
\newcommand{\ba}{\begin{eqnarray}}
\newcommand{\ea}{\end{eqnarray}}
\newcommand{\baa}{\begin{eqnarray*}}
\newcommand{\eaa}{\end{eqnarray*}}
\newcommand{\bb}{\begin{thebibliography}}
\newcommand{\eb}{\end{thebibliography}}
\newcommand{\ci}[1]{\cite{#1}}
\newcommand{\bi}[1]{\bibitem{#1}}
\newcommand{\lab}[1]{\label{#1}}

\begin{document}
\begin{center}
\begin{flushright}
JINR preprint E2-93-454.
\end{flushright}
\vspace{15mm}

{ \bf WHAT CAN BE LEARNT FROM THE NEW  UA4/2 DATA }\\[10mm]
 {\large O.V.Selyugin}\\
Bogolubov Laboratory of Theoretical Physics,\\
Joint Institute for Nuclear Research, Dubna\\
Head Post Office P.O.Box 79, 101000 Moscow, Russia$^{1} $\\[10mm]
\end{center}

\begin{abstract}
 A careful analysis of the new data of the UA4/2 collaboration reveals that
these data give an essentially large value of the $\rho = Re T(s,t)/Im(s,t)$
that does not contradict the early UA4 experiment.
There is the reason to think also that this experiment reveals for the first
time a real possibility of the existence of the spin-flip amplitude at
superhigh energies in the range of small transfer momenta.
\end{abstract}

\vspace{2cm}
\begin{flushleft}
------------\\
$^{1} $ E-mail: selugin@thsun1.jinr.dubna.su
\end{flushleft}

\newpage
\phantom{.}
\vspace{10mm}

      The elastic hadron-hadron scattering plays an important role in the
investigation of strong interactions. For the description of the interaction
at small distances we have the exact  theory, QCD, but for the interaction
at large distances, that is the basis for the elastic  scattering  at
small angles, the calculation in the framework of QCD is impossible at
present. These two domains are tightly connected and  the  experimental
determination of the parameters of elastic  scattering is very important
for the development of the  modern  strong interaction  theory \ci{land1}.

     The potential of interaction of charged hadrons is a sum of coulomb and
nuclear interactions. After the eikonal summation the terms with the coulomb
and nuclear interactions appear. As a result, the total interaction amplitude
has a complicated structure and depends on the spin parameters. However,
currently, at sufficiently high energies and small scattering angles the
contribution of spin-flip amplitudes can usually be neglected \ci{bl}.

   A surprisingly high value of the ratio $\rho$ of the real to imaginary
part of the forward elastic scattering amplitude obtained by the UA4
Collaboration \cite{ua4}  gave rise to various theoretical interpretations
\cite{odd1}. The new experiment was made by the UA4/2 Collaboration \cite{ua42}
to confirm or to specify this value of the $\rho$. This experiment gives
unique experimental data. In it a very small value of $|t|$ was reached
for a large enough energy and the differential cross section was  obtained
with sufficiently small errors. In a preliminary publication the authors
gave the calculated value $\rho = 0.135 \pm .015$. This value of $\rho$
refutes the previous UA4 data and is close to many odderon models.
But is it really so?

   In paper \ci{mart} the existence of four possibilities is noticed
for understanding the large value of $\rho$. In this work, we carry out
a careful analysis of the new experimental  UA4/2 data trying to take
into account only these experimental data. This analysis shows, from our
view point, that the value of $\rho$  is sufficiently large and has no
contradiction with the experimental UA4 data. Moreover, these data,
maybe, show for the first time a real possibility for the existence  of
the spin-flip amplitude at superhigh energies in the range of small $|t|$.

   The differential cross sections measured in the experiment
are described by the square of the scattering amplitude
\begin{eqnarray}
d\sigma /dt = \pi \ (F^2_C (t)+ (1 + \rho (s,t)) \ Im F^2_N(s,t)
 \mp 2 (\rho (s,t) +\alpha \varphi )) \ Im F_N F_C)   \lab{ds2}
\end{eqnarray}
where $F_C = \mp 2 \alpha G^2/|t|$ is the coulomb amplitude;
$\alpha$ is the fine-structure constant  and $G (t)$ is  the  proton
electromagnetic form factor squared;
$\rho(s,t) = Re\ F(s,t) / Im\ F(s,t)$.
Just this formula is used for the fit  of  experimental  data
determining the coulomb and hadron amplitudes and the coulomb-
hadron phase to obtain the value of $\rho(s,t)$. Solving (\ref{ds2}) for
the imaginary part of the hadron amplitude, we get
\begin{eqnarray}
Im\ F_N(s,t)=-\ \frac{\rho + \alpha \varphi}{ 1+\rho ^2} \ F_C+\lbrack
{(\rho +\alpha\varphi )^2 \over (1+\rho ^2)^2} F_C^2 +\ {1\over (1+\rho ^2)}
({1\over \pi} {d\sigma (s,t)\over dt} - F^2_C)\rbrack^{1/2}.   \lab{imro}
\end{eqnarray}
      Here, the one-to-one correspondence of the imaginary part  of
the hadron amplitude and $\rho (s,t)$ is seen. At each point of the
transfer momentum, using $\rho (s,t)$ we can obtain $Im F (s,t)$ from
the experimental data on the  differential  cross  sections.
The phase  of the coulomb-hadron interaction  has been calculated
and discussed  by many authors \ci{fi} and  has the form \ci{can}
\ba
\varphi (s,t)&=&\mp [\gamma +\ln (B|t| /2)+\ln (1 + 8/(B\Lambda ^2))
+\nonumber\\
             & & (4|t|/\Lambda ^2)\ \ln (4|t|/\Lambda^2) + 2|t|/\Lambda^2],
\lab{fit}
\ea
here $\Lambda$ is a constant entering into the dipole form factor.
The pure hadron amplitude is represented in the exponential form in the range
of the diffraction peak and a small interval of $t$:
\ba
F(s,t) =  A \ (i +\rho) \  \exp (- B(s,t)/2 \ |t|), \lab{fri}
\ea
$A$ is the interaction effective constant.
In the experiment the coefficient $\rho(s,t)$ is obtained from the analysis
of the differential cross sections in the region of the coulomb-hadron
interference where the coulomb and hadron amplitudes are nearly equal
to one another and their interference term has the maximum relative
contribution. The imaginary part of the amplitude of elastic scattering
is connected with the total cross section
$$\sigma_{tot}(s) = 4\pi Im\ T(s,t=0)$$ .

    In work \cite{ua42} the value of $\rho$ was obtained by using
formulae (\ref{ds2}), but the value of $A$ in (\ref{fri}) was determined from
another
experiment \ci{ua4s}. This experiment gives
$\sigma_{tot} \cdot (1+\rho^2)=63.3 mb.$,
and for $\rho=.15$ one obtains $\sigma_{tot}=61.9 mb$.
It is just the value used in work \ci{ua42} to compute $\rho$.
Therefore the formula for the imaginary part of the scattering amplitude
is represented as
\ba
Im T(s,t) = A_{\sigma} \cdot exp(-B \cdot |t|); \\  \nonumber
  A_{\sigma} = (\sigma_{tot}^1 \cdot (1+\rho_{1}^2)=63.3)/(1 + \rho_{2}^2)
/(4 \pi \cdot 0.38937966). \lab{as}
\ea
The constant $A_{\sigma}$ is in fact dependent on $\sigma_{tot}^1$
and $\rho_{1}$ defined from another experiment. Note that the error
of $\sigma_{tot}^1$ is not included in the final error of $\rho_2$.

  As is noted in  previous paper \cite{sel}, the procedure of
extrapolation of the imaginary part of scattering amplitude is very
significant for determining $\sigma_{tot}$.
     The importance of the extrapolated contribution is seen from
paper \ci{carb} where the contribution to $\sigma_{tot}$ of
$\sigma_{obs}$, the directly measured value, and of $\Delta \sigma_{el}$ and
$\Delta \sigma_{inel}$, the extrapolated contributions of the elastic and
inelastic cross sections, are shown at energies $\sqrt s = 30.6\ GeV,
52.8\ GeV$ and $62.7\ GeV$. One can see that the growth of the total cross
sections is due to $\Delta \sigma_{el}$ by $50\%$ for $p p$ and
nearly by $100\%$ for  $p \bar p$ scattering.

   If we can determine the value of $\rho$ using (\ref{imro}), then
we obtain almost the same value (see Table 1, variant 1) $\rho=0.137 \pm .007$,
the error is only statistical. Insignificant difference from the result
\ci{ua42} may consist in more precise numerical calculations.
Let us take the value $A_{\sigma} \rightarrow A$ as a free parameter.
In this case we obtain $\rho= 0.148 \pm 0.018$
(see var. 2 in Table 1).

   In these two variants we suppose that the amplitude has a constant slope
in this range of transfer momenta. Let us examine this supposition as
this unique experiment allows us to do it. We will  reduce the number of
considered  experimental points from 99,95,90,85 ... to 50 and therefore
the interval of transfer momenta from $|t|=120. \cdot 10^{-3} GeV^2$  to
$|t|=18 \cdot 10^{-3} GeV^2$  and will  obtain a new value of $\rho_{i}$
and $B_{i}$.  We show that the value of $\rho_{i}$ grows and the value
of $B_{i}$ decreases (see fig. 1 and 2  or Table II).
Therefore our method of the determination of $\rho$ depends on the
investigated interval of $|t|$ .

     Let us examine another form of the scattering amplitude which
is $|t|$ -dependent in form (see var. 3,4 and 5,6 in Table 1).
For variants 3,4 we also take the constant $A_{\sigma}$
as in work \ci{ua42} and obtain some decrease of $\chi^2$ and growth of $\rho$.
The values of the constant $C$ are $0.86\pm0.48$ and $-0.15 \pm 0.08$
respectively. In variants 5,6 we change again $A_\sigma to A$
as a free parameter. The $\chi^2$ continues to decrease and $\rho$ grows.
In these variants the values of the constant $C$ are
$1.80\pm0.56$ and $-0.27 \pm 0.097$ respectively. We obtain the decrease
of $\chi^2$ almost by $8\%$ and large growth of $\rho$. But which form of
the scattering amplitude will be obtained in these cases? As the value of
the coefficient $C$ is positive in variants 3,5 and negative in
variants 4,6, we obtain the decrease of the slope of the scattering amplitude
in these cases when $t \rightarrow 0$. It is to be recalled that the slope
of differential cross sections grows in the range of $|t|$ near
$0.05 - 0.4 GeV^2$ and now we see that it decreases when
$|t| \rightarrow 0$. This is very unusual and imposes strong restrictions
both on the ordinary pomeron and the odderon models.
This behavior of the scattering amplitude is, maybe, due to some oscillations
of it \ci{car} or can be obtained by taking into
account the next rescattering term of the amplitude. In the latter case we
also obtain a large value of $\rho$ (see v. 7,8 in Table 1). This requires
to include one or two additional free parameters and raises problems
with the summation of non-leading terms of the scattering amplitude.
This leads us to the range of theoretical models whereas we wish to stay
only in the framework of this experiment.

     However, maybe, the matter is simpler.
Let us regard the possibility of the contribution of the spin-flip
amplitude to the differential cross sections. The simplest form of this
amplitude that gives a sufficiently large contribution in the range of
small $|t|$ and does not change the form of the differential cross sections
at large $|t|$ is, for example, as follows:
\ba
F^{+-}(s,t) = \sqrt{|t|} \cdot A \cdot exp(- B \cdot |t|).
\ea
In this case we don't introduce additional free parameters. As we can see
>from variant 9 of Table 1, we obtain the same minimum of $\chi^2$  without
additional parameters for the slope. Let us examine again the behavior of
our parameters as a function of the regarded interval of transfer momenta.
We obtain that in this variant the values of the slope and $\rho$ do not
change with decreasing intervals of $|t|$ (see fig. 1 and 2  or Table II).
This shows that the possibility of the existence of the spin-flip
amplitude and its manifestation in this experiment is sufficiently large.
However, we obtain a very large value of $\sigma_{tot} \cdot (1+\rho^2)$,
different from $63.3\pm1.5mb$ by three errors. The degree of
the increase of $\sigma_{tot}$ is examined \ci{land2}. It is clear
that such a large value of $\sigma_{tot}$ requires special explanation.
If we use the fixed value of $A_\sigma$ and make $A_{spin}$ the free parameter
then we obtain variant 10. The increase by one error for $\sigma_{tot}$
leads to $A_{\sigma2}$ in variant 11. Evidently, there is a direct
relationship between the values of $\rho$ and $A_\sigma$ .

       Thus, we can make the following conclusion. The new UA4/2 experimental
data measured with very small errors and in a sufficiently small
interval of transfer momenta allow us to calculate the normalization
coefficient, determine the values of $\rho$ and the slope - $B$ based only
on this experiment. The analysis of these experimental data gives an
essentially large value of $\rho $, most likely, $\rho= 0.19\pm0.03$
(only statistical error).
This contradicts nether the value $\rho=0.168\pm0.018 $, when we lean
upon the early obtained $\sigma_{tot}$, nor $\rho=0.24\pm.045$,
when we take $\sigma_{tot}$ as a free parameter.
The question of manifestation of the spin-flip amplitude in the diffraction
scattering is exceptionally interesting. We show that this possibility is
sufficiently probable. This is tightly connected with the value of
$\sigma_{tot}$.It would be very important to have some experimental points in
the range before $|t|_{max}$ at which the relative maximum of
interference of the coulomb  nucleon amplitudes occurs. In this case
the normalization will be entirely determined by the coulomb amplitude.
It sharply decreases the errors of the obtained $\sigma_{tot}$, $\rho$ and
$B$. The manifestation  of spin-flip amplitude requires
polarization experiments in the diffraction range.
Some models predict sufficiently large effects in this energy range
(see \ci{sof,gol}) especially in the range of the diffraction minimum for
the polarization and in the range of $|t|=1 \div 3 GeV^2$ for $A_{NN}$.

     {\it Acknowledgement.} {\hspace {0.5cm} The author expresses his deep
gratitude to D.V.Shirkov, A.N. Sissakyan for support in this work and
to L. Jenkovsky,  V.A. Meshcheryakov, S.V. Goloskokov for
fruitful discussions of the problems considered in this paper.

\newpage
\phantom{.}
\vspace{.5cm}

{\bf TABLE 1}
\vspace{.5cm}

\begin{tabular}{|c|c|c|c|c|c|} \hline
N& $F(s,t)^{++}$ & $ \sum\limits_{i=1}^{99} \chi^2_i$ &$B$ $(GeV^{-2})$&
$\rho $ &$ \sigma_{tot}$ $mb$ \\ \hline
& & & & &  \\
1&$A_{\sigma} \cdot exp(-B/2 \cdot |t|)$&106.52&$15.52\pm0.06$&$.137\pm.007$
&62.13 \\
&  & & & & \\
2&$A \cdot  exp(-B/2 \cdot |t|)$         &106.06&$15.50\pm0.07$&$.148\pm.018
$&62.79  \\
&  & & & & \\
3 &$A_{\sigma} \cdot exp(-B/2 \cdot |t|-C*t^2)$&103.24&$15.16\pm0.20$
&$.147\pm.009$&61.96    \\
&  & & & & \\
4 &$A_{\sigma} \cdot exp(-B/2 \cdot |t|-C \cdot \sqrt{|t|})$&102.90
&$16.21\pm0.36$&$.168\pm.018$& 61.56  \\
&  & & & & \\
5&$A \cdot  exp(-B/2 \cdot |t|-C \cdot t^2)$   &100.20&$14.91\pm0.25$
&$.188\pm.027$&63.74 \\
&  & & & & \\
6&$A \cdot  exp(-B/2 \cdot |t|-C \cdot \sqrt{|t|})$&98.44&$16.66\pm0.43$
&$.2437\pm.045$&63.4  \\
&  & & &  &\\
7&$ A_1 \cdot  exp(-B/2 \cdot |t|)$   &99.42&$16.76\pm0.43$&$.197\pm.029$
&63.89  \\
 &$ -A_2 \cdot  exp(-B\cdot |t|)$& & & &             \\
&  & & &  &\\
8&$A_1 \cdot  exp(-B_1/2 \cdot |t|)$  &98.0&$15.74\pm0.26$&$.236\pm.061$&64.26
\\
 &$ -A_2 \cdot  exp(-B_2/2 \cdot |t|)$& & & &             \\
&  & & &  &\\
9&$A \cdot  exp(-B/2 \cdot |t|)$ and
&98.62&$15.67\pm0.065$&$.233\pm.022$&62.79  \\
 &$F^{+-}=\sqrt{|t|} \cdot  A \cdot  exp(-B\cdot |t|)$& & &  &            \\
&  & & &  &\\
10&$A_\sigma \cdot  exp(-B/2 \cdot |t|)$ and
&102.90&$15.63\pm0.08$&$.152\pm.011$&61.87  \\
 &$F^{+-}=\sqrt{|t|} \cdot  A_s \cdot  exp(-B\cdot |t|)$& & &  &            \\
&  & & &  &\\
11&$A_{\sigma2} \cdot  exp(-B/2 \cdot |t|)$ and
&99.8&$15.64\pm0.08$&$.178\pm.011$&62.82  \\
 &$F^{+-}=\sqrt{|t|} \cdot  A_s \cdot  exp(-B\cdot |t|)$& & &  &  \\  \hline
\end{tabular}
\newpage

{\bf TABLE II}
\vspace{1.5cm}

\begin{tabular}{|c|c|c|c|c|c|} \hline

$\sum N_i$ &  $ <|t|_{up} $ &
\multicolumn{2}{c|}{$ B ( GeV^2) $}
& \multicolumn{2}{c|}{$ \rho$} \\ \cline{3-6}

&$10^{-3}GeV^2$& variant 2& variant 9    & variant 2    & variant 9 \\ \hline
99 & 120 &$15.50\pm.06$  & $15.67\pm.06$ & $.148\pm.018$ &$ .233\pm.021$   \\
95 & 110 &$15.53\pm.07$  & $15.73\pm.07$ & $.144\pm.019$ &$ .223\pm.022$  \\
90 & 97.5&$15.45\pm.07$  & $15.70\pm.07$ & $.154\pm.020$ &$ .227\pm.023$
\\
85 & 85.0&$15.39\pm.08$  & $15.70\pm.08$ & $.161\pm.021$ &$ .226\pm.023$    \\
80 & 72.5&$15.28\pm.09$  & $15.67\pm.10$ & $.173\pm.022$ &$ .232\pm.024$  \\
75 & 60.0&$15.26\pm.13$  & $15.75\pm.13$ & $.175\pm.023$ &$ .224\pm.025$
\\
70 & 47.5&$15.12\pm.17$  & $15.75\pm.18$ & $.186\pm.025$ &$ .223\pm.028$    \\
65 & 38.0&$14.93\pm.23$  & $15.69\pm.25$ & $.200\pm.027$ &$ .226\pm.031$  \\
60 & 32.0&$14.81\pm.31$  & $15.66\pm.32$ & $.205\pm.030$ &$ .227\pm.033$
\\
55 & 28.0&$14.34\pm.42$  & $15.29\pm.43$ & $.230\pm.033$ &$ .247\pm.037$
\\
50 & 23.0&$14.77\pm.64$  & $15.82\pm.63$ & $.211\pm.039$ &$ .224\pm.040$
\\
  &  & & &  & \\ \hline
\end{tabular}
\phantom{.}
\vspace{2cm}

\begin{center}
{\large                         Figure captions     } \\
\end{center}

Fig.1.  The dependence of the $\rho$ with the examined interval of $|t|$; \\
        $\times\!\!\!\!\!\mid$ - for the variant 2 ,
	$\circ\!\!\!\!\mid$ -for the variant 9 (see Table 1).

Fig.2.  The dependence of the slope - $B$ with  the examined interval
        of $|t|$ ; \\
        $\times\!\!\!\!\!\mid$ - for the variant 2 ,
	$\circ\!\!\!\!\mid$ -for the variant 9  (see Table 1).
\newpage


\end{document}